\documentclass[a4paper,11pt]{article}
\pdfoutput=1

\usepackage{jheppub}

\usepackage{subfig}
\usepackage[utf8]{inputenc} % if needed

\newcommand{\action}{\mathcal{S}}
\newcommand{\pathdiff}[1]{\!\mathcal{D}#1\,}
\newcommand{\diff}[1]{\!\mathrm{d}#1\,}

\newcommand{\braket}[2]{\left<#1|#2\right>}
\newcommand{\nn}{\nonumber\\}

\newcommand{\f}[1]{\mbox{\boldmath$#1$}}

\newcommand{\bea}{\begin{eqnarray}}
\newcommand{\ea}{\end{eqnarray}}
\newcommand{\eea}{\end{eqnarray}}
\newcommand{\ord}{\,{\cal O}}

\title{Dynamically assisted Sauter-Schwinger effect
in inhomogeneous electric fields}

\author{Christian Schneider}
\emailAdd{christian.schneider@uni-due.de}
\author{and Ralf Sch\"utzhold}
\emailAdd{ralf.schuetzhold@uni-due.de}
\affiliation{Fakult\"at f\"ur Physik, Universit\"at Duisburg-Essen,
Lotharstrasse 1, 47057 Duisburg, Germany}

\abstract{%
Via the world-line instanton method, we study electron-positron pair creation
by a strong (but sub-critical) electric field of the profile $E/\cosh^2(kx)$ 
superimposed by a weaker pulse $E'/\cosh^2(\omega t)$.
If the temporal Keldysh parameter $\gamma_\omega=m\omega/(qE)$ exceeds a
threshold value $\gamma_\omega^{\rm crit}$ which depends on the spatial Keldysh
parameter $\gamma_k=mk/(qE)$, we find a drastic enhancement of the pair
creation probability -- reporting on what we believe to be the first analytic
non-perturbative result for the interplay between temporal and spatial field
dependences $E(t,x)$ in the Sauter-Schwinger effect.
Finally, we speculate whether an analogous effect
(drastic enhancement of tunneling probability) could occur in other
scenarios such as stimulated nuclear decay, for example.%
}

\begin{document} 
\maketitle
\flushbottom

%%%%%%%%%%%%%%%%%%%%%%%%%%%%%%%%%%%%%%%%%%%%%%%%%%%%%%%%%%%%%%%%%%%%%%%%%%%%%%%
\section{Introduction}
%%%%%%%%%%%%%%%%%%%%%%%%%%%%%%%%%%%%%%%%%%%%%%%%%%%%%%%%%%%%%%%%%%%%%%%%%%%%%%%

Despite the tremendous progress of quantum field theory as a fundamental
description of nature, our understanding of its non-perturbative properties
is still disappointingly incomplete.
In quantum electrodynamics (QED), for example, a striking non-perturbative
phenomenon is the Sauter-Schwinger effect predicting the creation of
electron-positron pairs out of the vacuum by a strong electric field
\cite{Sauter,Euler,Weisskopf,Schwinger}.
In case of a constant (sub-critical) electric field $E<E_{\rm crit}$, 
the pair creation probability behaves as  ($\hbar=c=1$)
\bea
\label{constant}
P_{e^+e^-}
\sim
\exp\left\{-\pi\,\frac{m^2}{qE}\right\}
=
\exp\left\{-\pi\,\frac{E_{\rm crit}}{E}\right\}
\,,
\ea
where $E_{\rm crit}=m^2/q\approx 1.3\times10^{18}\, \textrm{V}/\textrm{m}$
denotes the Schwinger critical field.
Unfortunately, the dependence of this non-perturbative phenomenon on the
field profile $\f{E}(t,\f{r})$ away from the constant field approximation
is still mostly \textit{terra incognita}.
There are many results for fields which depend on one coordinate only,
such as space $x$ or time $t$ 
(see, e.g., \cite{Brezin+Itzykson,Narozhnyi+Nikishov,Popov-71,Kennedy,
Dietrich,Gies+Klingmuller,Kim+Page-07,Kleinert,Hebenstreit-sub-cycle,Stokes,
Dumlu,Kim+Schubert,Dumlu+Dunne,Akkermans+Dunne,Strobel+Xue}),
one of the light-cone coordinates $x_\pm=t\pm x$ (see, e.g.,
\cite{Nikishov+Ritus-1,Nikishov+Ritus-2,Woodard,Fried, Hebenstreit-lightfront,
Torgrimsson,Ilderton}), or other linear combinations of $x$ and $t$
\cite{interpolating}.
In these cases, the underlying (Dirac or Klein-Fock-Gordon) equation
simplifies to an ordinary differential equation 
(allowing for a WKB approach, for example, see also 
\cite{Popov-72,Popov+Marinov,Popov-73,Kluger}). 

However, to the best of our knowledge, there are no analytic non-perturbative 
results for fields $E(t,x)$ which genuinely depend on space $x$ and time $t$.
So far, this case has only been treated numerically via the
Wigner formalism
(see, e.g., 
\cite{Hebenstreit-quantum-kinetic,Hebenstreit-spacetime,Hebenstreit-Wigner})  
or a direct integration of the Dirac equation 
(see, e.g., \cite{Keitel,Jiang-11,Jiang-12}).  
This lack of understanding is not only unsatisfactory from a theoretical point
of view.
A deeper insight into the impact of space-time dependent fields is also 
highly desirable in view of experimental efforts with lasers 
\cite{Bunkin+Tugov,Ritus,Nikishov,Piazza,Bulanov-04,Narozhny,Bulanov-06,
Mourou,Dunne-overview-09,Attainability,Labun+Rafelski}, for example,  
aiming at a verification of this non-perturbative pair-creation effect.
\footnote{
Apart from the laser laboratories BELLA (Berkley, USA) and VULCAN (Oxford, UK),
which are approaching the strong-field (non-linear) QED regime, we would like 
to mention the European ELI program, the Russian XCELS initiative, 
or the Chinese SIOM facility, for example.}
In the following, we venture a first step into this direction and employ
the world-line instanton technique 
(see, e.g., \cite{Feynman,Afflek,Kim+Page-02,Dunne+Schubert,
Dunne,Kim+Page-06,Dunne-overview-08,Dumlu+Dunne-complex,
Schubert-Lectures}) 
in order to study the superposition of a spatial and a temporal field
pulse as an example for a genuinely space-time dependent field.

%%%%%%%%%%%%%%%%%%%%%%%%%%%%%%%%%%%%%%%%%%%%%%%%%%%%%%%%%%%%%%%%%%%%%%%%%%%%%%%
\section{World-line instanton method}
%%%%%%%%%%%%%%%%%%%%%%%%%%%%%%%%%%%%%%%%%%%%%%%%%%%%%%%%%%%%%%%%%%%%%%%%%%%%%%%

Let us start with a brief review of the world-line instanton method,
see, e.g., \cite{Feynman,Afflek,Kim+Page-02,Dunne+Schubert,
Dunne,Kim+Page-06,Dunne-overview-08,Dumlu+Dunne-complex, 
Schubert-Lectures}. 
Since the electron spin does not affect the exponent of the pair creation
probability \cite{Dunne+Schubert}, 
we consider the vacuum persistence amplitude of scalar QED
\begin{equation}
\braket{0_\text{out}}{0_\text{in}} 
=
\iint\pathdiff{\phi}\,\pathdiff{\phi^*}\,
      e^{i \int\diff{^4x \left( |D_\mu\phi|^2 - m^2|\phi|^2 \right)}}
\,,
\end{equation}
with the covariant derivative $D_\mu = \partial_\mu + i q A_\mu$.
After analytic continuation to Euclidean space, this functional path integral
can be translated into the world-line representation \cite{Feynman} 
where $\pathdiff{\phi}\,\pathdiff{\phi^*}$ is replaced by the sum over all 
closed loops $x_\mu(s)$ in Euclidean space.
Then, via the saddle point method (with the electron mass $m$ playing the
role of the large expansion parameter), the pair creation probability can be
estimated as
\bea
\label{probability}
P_{e^+e^-}
=
1-|\braket{0_\text{out}}{0_\text{in}}|^2
\sim
e^{-\action}
\,,
\ea
with the  world-line instanton action 
\footnote{
Although one could also consider complex instantons, $x^\mu(s)$ is purely 
real in our case while $\hat A_\mu$ is purely imaginary, 
cf.~Eq.~(\ref{Euclidean-vector-potential}).   
} 
\bea
\label{instanton-action}
\action =
ma +
i q \int\limits_0^1\diff{s}\,\dot{x}^\mu A_\mu(x^\nu)
\,.
\ea
Here $\dot{x}_\mu=\mathrm{d}x_\mu/\mathrm{d}s$ denotes
the proper-time derivative of a closed $x_\mu(s=0)=x_\mu(s=1)$ world-line loop
$x_\mu(s)$ as a solution of the instanton equations
\begin{equation}
\label{instanton}
m \ddot{x}_\mu = i q F_{\mu\nu} \dot{x}^\nu a
\end{equation}
with $\ddot{x}_\mu=\mathrm{d}^2x_\mu/\mathrm{d}s^2$ and
$\dot{x}_\nu\dot{x}^\nu=a^2=\rm const$.

%%%%%%%%%%%%%%%%%%%%%%%%%%%%%%%%%%%%%%%%%%%%%%%%%%%%%%%%%%%%%%%%%%%%%%%%%%%%%%%
\section{Sum of Sauter pulses}
%%%%%%%%%%%%%%%%%%%%%%%%%%%%%%%%%%%%%%%%%%%%%%%%%%%%%%%%%%%%%%%%%%%%%%%%%%%%%%%

Now let us apply the world-line instanton method to a space-time
dependent electric field
\bea
\label{Sauter}
\f{E}(t,x)
=
\left(
\frac{E}{\cosh^2(kx)}+\frac{E'}{\cosh^2(\omega t)}
\right)
\f{e}_x
\ea
consisting of a strong spatial Sauter \cite{Sauter} pulse $\propto E$
and a weaker temporal Sauter pulse $\propto E'$ where both field
strengths are sub-critical $E'\ll E \ll E_{\rm crit}=m^2/q$.
Furthermore, in order to be in the non-perturbative regime, we assume slowly
varying pulses $\omega,k\ll m$.
For convenience, we introduce the spatial and temporal Keldysh \cite{Keldysh}
parameters via 
\bea
\label{Keldysh}
\gamma_k=\frac{mk}{qE}
\,,\;
\gamma_\omega=\frac{m\omega}{qE}
\,.
\ea
It will be most convenient to represent the spatial pulse by the scalar 
potential $A_0(x)$ but the temporal pulse by the vector potential $A_1(t)$.
Then, after Wick rotation, the Euclidean vector potential reads
\bea
\label{Euclidean-vector-potential}
A_0(x_1)=i\,\frac{E}{k}\,\tanh(kx_1)
\,,\;
A_1(x_0)=i\,\frac{E'}{\omega}\,\tan(\omega x_0)
\,,
\ea
with $x_0=it$ and $x_1=x$ as well as $A_2=A_3=0$.
As a result, the instanton equations~(\ref{instanton})
assume the form
\bea
\label{instanton-cosh}
\ddot x_0
&=&
+\frac{qEa}{m}
\left(
\frac{1}{\cosh^2(kx_1)}-\frac{E'}{E}\,\frac{1}{\cos^2(\omega x_0)}
\right)
\dot x_1
\,,
\nn
\ddot x_1
&=&
-\frac{qEa}{m}
\left(
\frac{1}{\cosh^2(kx_1)}-\frac{E'}{E}\,\frac{1}{\cos^2(\omega x_0)}
\right)
\dot x_0
\,,
\ea
and are analogous to the planar motion of a charged particle in a magnetic
field $\f{B}(\f{r})=B(x,y)\f{e}_z$.

\begin{figure}
\begin{center}
\subfloat[][$\omega<\omega^{\rm crit}$]
{\includegraphics[width=0.33\linewidth]{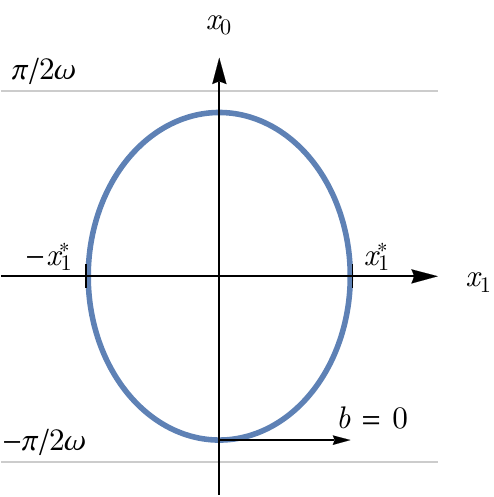}}
\hspace{0.2\textwidth}
\subfloat[][$\omega>\omega^{\rm crit}$]
{\includegraphics[width=0.33\linewidth]{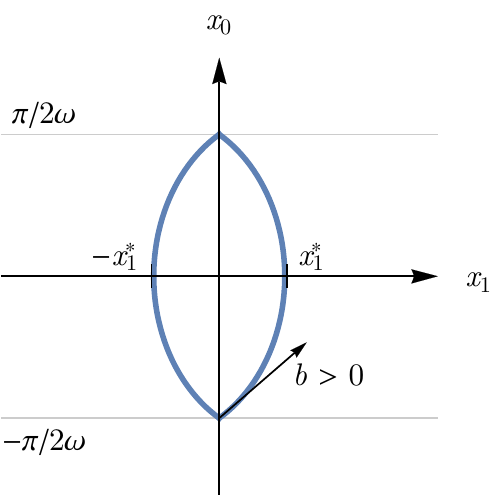}}
\end{center}
\caption{Sketch of the instanton trajectories in the Euclidean 
$x_0,x_1$-plane: 
Left (a): for frequencies below threshold  
$\omega<\omega^{\rm crit}\leftrightarrow\gamma_\omega<\gamma_\omega^{\rm crit}$
with $\gamma_\omega^{\rm crit}$ being given by Eq.~(\ref{threshold}), 
the trajectory does not 
reach the poles at $\pm\pi/(2\omega)$ and thus it has nearly the same 
form as in the absence of the temporal pulse $\propto E'$. 
Right (b): for frequencies above threshold  
$\omega>\omega^{\rm crit}\leftrightarrow
\gamma_\omega>\gamma_\omega^{\rm crit}$,
the instanton trajectory is effectively reflected at the poles.
}
\label{trajectories}
\end{figure}

Due to $E'/E\ll1$, the second term is negligible unless
$\cos^2(\omega x_0)$ becomes very small -- which happens near the poles 
of $E(x_0,x_1)$ at $\omega x_0=\pm\pi/2$.
Away from these poles, we may omit the second term and the above equations
can be integrated approximately to 
\bea
\dot x_0
&=&
\frac{a}{\gamma_k}\,\tanh(kx_1)+ab
\,,
\nn
\dot x_1
&=& \pm a \sqrt{1-\left(\frac{\tanh(kx_1)}{\gamma_k}+b\right)^2}
\,.
\ea
As mentioned after Eq.~(\ref{instanton}), the constant $a$ is given by
$\dot{x}_\nu\dot{x}^\nu=a^2=\rm const$.
The other  integration constant $b$ determines the velocity $\dot x_0$ just
before (or just after) crossing the $x_0$-axis, see Fig~\ref{trajectories}. 

Near the poles $\omega x_0\approx\pm\pi/2$, on the other hand, the second
term becomes important.
Similar to the reflection of a charged particle at the region of a very
strong magnetic field, the instanton trajectory is basically reflected
by the ``wall'' at $\omega x_0\approx\pm\pi/2$ if it reaches
out far enough.
Since this reflection occurs during a very short proper time $\Delta s$,
we may neglect the regular terms in Eq.~(\ref{instanton-cosh}) and keep
only the divergent contributions.
Then, the equation for $x_1$ can be integrated approximately to
\bea
\dot x_1\approx\frac{qE'a}{m\omega}\,\tan(\omega x_0)+\dot x_1^{\rm in}
\,,
\ea
and thus the equation for $x_0$ becomes
\bea
\label{approximation-poles}
\ddot x_0
\approx-
\frac{(qE'a)^2}{m^2\omega}\,
\frac{\tan(\omega x_0)}{\cos^2(\omega x_0)}
\sim
\frac{1}{(\omega x_0\pm\pi/2)^3}
\,.
\ea
As a result, the perpendicular velocity $\dot x_0$ is reversed by that
reflection while the parallel velocity $\dot x_1$ has the same value
$\dot x_1^{\rm in}$ before and after the reflection, 
see Fig~\ref{trajectories}. 

%%%%%%%%%%%%%%%%%%%%%%%%%%%%%%%%%%%%%%%%%%%%%%%%%%%%%%%%%%%%%%%%%%%%%%%%%%%%%%%
%%%%%%%%%%%%%%%%%%%%%%%%%%%%%%%%%%%%%%%%%%%%%%%%%%%%%%%%%%%%%%%%%%%%%%%%%%%%%%%

%%%%%%%%%%%%%%%%%%%%%%%%%%%%%%%%%%%%%%%%%%%%%%%%%%%%%%%%%%%%%%%%%%%%%%%%%%%%%%%
\section{Tunneling probability}
%%%%%%%%%%%%%%%%%%%%%%%%%%%%%%%%%%%%%%%%%%%%%%%%%%%%%%%%%%%%%%%%%%%%%%%%%%%%%%%

From Eq.~(\ref{approximation-poles}), we may estimate the contribution to the 
instanton action~(\ref{instanton-action}) stemming from the short proper-time 
period of the reflection and obtain that it is suppressed as 
$(E'/E)\ln(E'/E)$ for small $E'$. 
Thus, for $E\gg E'$, we may approximate the instanton action by the part
between the reflections 
\bea
\action
\approx
m a - \frac{qE}{k} \int\limits_0^1 \diff{s}\,\tanh(kx_1)\,\dot{x}_0
\,.
\ea
In order to calculate the above integral, we split the closed loop in 
Fig.~\ref{trajectories}(b) into four quarters: 
starting at the lower reflection point $x_0=-\pi/(2\omega)$ with $x_1=0$, 
we go to the spatial turning point $x_1^*$ (where $x_0=0$),  
then from there to the upper reflection point $x_0=+\pi/(2\omega)$ 
with $x_1=0$, 
and then to $-x_1^*$ (where $x_0=0$), 
and finally back to $x_1=0$  and $x_0=-\pi/(2\omega)$. 
Since each quarter yields the same contribution, we get
\bea
\label{action}
\action
\approx
ma - \frac{4 m}{\gamma_k}
\int\limits_0^{x_1^*}\diff{x_1}\,
\frac{\tanh(k x_1)\left(
\tanh(k x_1) + \gamma_kb\right)}
{\sqrt{\gamma_k^2-\left(
\tanh(kx_1)+ \gamma_kb\right)^2}}
\,,\quad
\ea
where $x_1^*$ denotes the spatial turning point given by
\bea
\tanh(kx_1^*)+\gamma_kb=\gamma_k
\,,
\ea
i.e., the zero of the square root in the integral in Eq.~(\ref{action})
where $\textrm{d}x_1/\textrm{d}x_0=0$.
The constant $a$ is determined by $\dot{x}_\nu\dot{x}^\nu=a^2$
and $x_\mu(s=0)=x_\mu(s=1)$ which gives
\bea
a=
\frac{4}{\gamma_k}
\int\limits_0^{x_1^*}
\frac{\diff{x_1}}
{\sqrt{\gamma_k^2-\left(\tanh(kx_1)+ \gamma_kb\right)^2}}
\,.
\ea
The remaining integration constant $b$ depends on the frequency $\omega$.
If $\omega$ is too small and thus the poles at $\omega x_0=\pm\pi/2$
are too far away, the instanton trajectory is not reflected at all and
thus we have $b=0$, see Fig~\ref{trajectories}. 
In case of reflection, the integration constant $b$ is non-zero and determined
by the implicit condition
\bea
\label{thresholdIntegral}
\frac{4 m}{\gamma_k}
\int\limits_0^{x_1^*}\diff{x_1}\,
\frac{\tanh(k x_1) + \gamma_k b}
{\sqrt{\gamma_k^2-\left(\tanh(kx_1)+ \gamma_k b\right)^2}}
=\frac{\pi}{2\omega}
\,.
\ea
Together with the above equations for $x_1^*$, $a$, and $b$,
Eq.~(\ref{action}) is the main result of this paper.

The threshold condition $b=0$ translates into
\bea
\label{threshold}
\gamma_\omega=\frac{\pi}{2}\,
\frac{\gamma_k\sqrt{1-\gamma_k^2}}{\arcsin(\gamma_k)}
\stackrel{{\rm Def}}{=}\gamma_\omega^{\rm crit}
\,.
\ea
Again, if the frequency is too low $\gamma_\omega<\gamma_\omega^{\rm crit}$,
the instanton trajectory is basically not affected by the poles at
$\omega x_0=\pm\pi/2$ leading to $b=0$ and thus the weak temporal pulse
$\propto E'$ has negligible impact.
In this case $b=0$, we get $x_1^*={\rm artanh}(\gamma_k)/k$ and all the
integrals can be carried our analytically, yielding the same results as
for a static Sauter pulse, see, e.g., \cite{Dunne+Schubert}.  
If the frequency exceeds this threshold value
$\gamma_\omega>\gamma_\omega^{\rm crit}$, on the other hand, the instanton
trajectory is reflected at the poles (i.e., $b>0$) and thus the instanton
action~(\ref{action}) is reduced by the weak temporal pulse $\propto E'$,
leading to a significant enhancement of the pair creation probability.
This enhancement is an example of the dynamically assisted Sauter-Schwinger 
effect, see 
\cite{assisted,catalysis,momentum-dependence,Orthaber,Monin+Voloshin,
Kleinert+Xue}, for an inhomogeneous electric field. 
In the homogeneous limit $\gamma_k\downarrow0$, the
threshold value~(\ref{threshold}) approaches $\gamma_\omega^{\rm crit}=\pi/2$
consistent with the results of \cite{assisted}. 
For $\gamma_k\uparrow1$, the threshold $\gamma_\omega^{\rm crit}$ scales
as $\gamma_\omega^{\rm crit}\approx\sqrt{1-\gamma_k^2}$, i.e., very small
frequencies $\omega$ can have a significant impact in this case.

Unfortunately, due to the implicit nature of the condition for $b$,
we cannot provide a closed analytical expression for $\action$.
However, near but above threshold, we can expand the involved quantities
and obtain the following approximate formula for the instanton
action:
\begin{equation}
\action=\frac{m^2}{qE}
\begin{cases}
\displaystyle
\frac{2\pi}{1+\sqrt{1-\gamma_k^2}}, & \text{for }\gamma_\omega\leq\gamma_\omega^{\rm crit}, \\
\begin{aligned}[b]
\frac{2\pi}{1+\sqrt{1-\gamma_k^2}}
-\pi\frac{(1-\gamma_k^2)^{3/2}}{\gamma_k^2(\gamma_\omega^{\rm crit})^4}
&\left[\gamma_\omega-\gamma_\omega^{\rm crit}\right]^2\\[-1em]
&+\ord\left(\left[\gamma_\omega-\gamma_\omega^{\rm crit}\right]^3\right), 
\end{aligned}& \text{for }\gamma_\omega > \gamma_\omega^{\rm crit}.
\end{cases}
\end{equation}
The first line (valid below and at threshold $\gamma_\omega\leq\gamma_\omega^{\rm crit}$) 
is just the result in the static case (see, e.g., \cite{Dunne+Schubert}).
For $\gamma_k=0$, we recover Schwinger's result
\cite{Schwinger} for a constant field~(\ref{constant}).
Above threshold $\gamma_\omega>\gamma_\omega^{\rm crit}$ on the other hand,
the action is reduced by the second-order term
$\propto[\gamma_\omega-\gamma_\omega^{\rm crit}]^2$.

%%%%%%%%%%%%%%%%%%%%%%%%%%%%%%%%%%%%%%%%%%%%%%%%%%%%%%%%%%%%%%%%%%%%%%%%%%%%%%%
\section{Numerical evaluation}
%%%%%%%%%%%%%%%%%%%%%%%%%%%%%%%%%%%%%%%%%%%%%%%%%%%%%%%%%%%%%%%%%%%%%%%%%%%%%%%

\begin{figure}
\begin{center}
\includegraphics{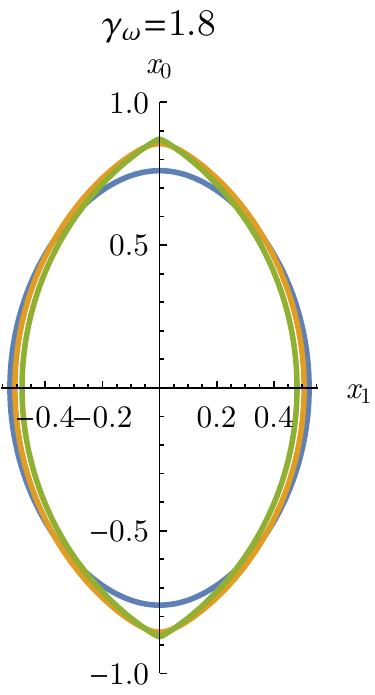}
\hspace{0.1\linewidth}
\includegraphics{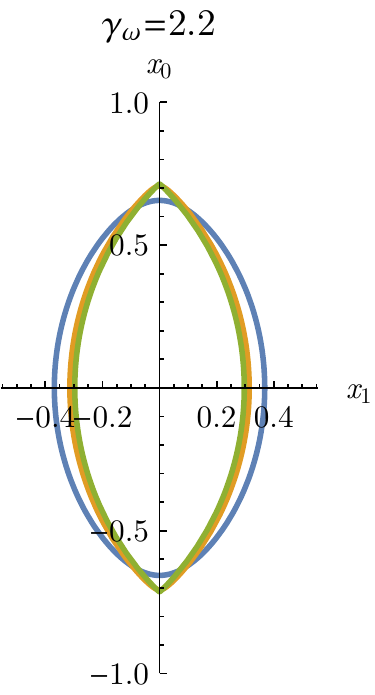}
\hspace{0.1\linewidth}
\includegraphics{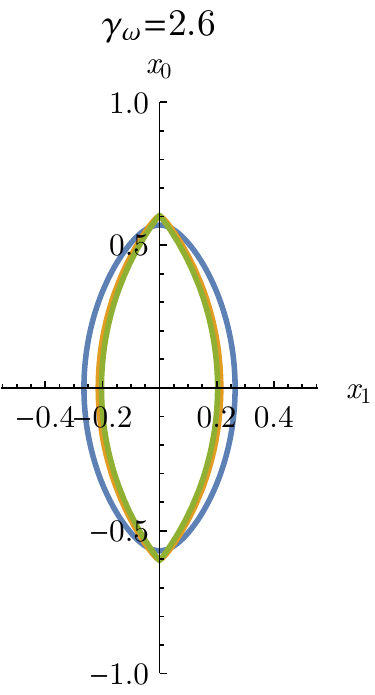}
\end{center}
\caption{
Instanton trajectories [in units of $m/(qE)$] found by numerically
integrating~\eqref{instanton-cosh} for different values of
$\gamma_\omega$ and $\gamma_k=0.5$ in all cases. 
The colors represent
$E'/E=10^{-1}$ (blue), $E'/E=10^{-2}$ (orange), and
$E'/E=10^{-3}$ (green). 
The limit $E'/E\to 0$ used in the analytic approximation 
(cf.~Fig.~\ref{trajectories}) 
is indistinguishable from the trajectory
with $E'/E=10^{-3}$.
}
\label{trajectoriesNumerical}
\end{figure}

Let us now compare the analytic approximation above to a numerical
solution of the full instanton equations \eqref{instanton-cosh}. 
To fulfill the periodicity constraints, a shooting method can be used: 
we numerically integrate \eqref{instanton-cosh}, varying the initial
conditions and $a$ until a closed solution is found using a root finding
algorithm.
Figure~\ref{trajectoriesNumerical} shows these closed solutions for
different values of $\gamma_\omega$ and $E'/E$ with a fixed
$\gamma_k=0.5$. 
As explained above, for small $E'$ the time dependent pulse acts like 
a reflecting barrier. 
For finite values of $E'$ such as $E'/E=10^{-1}$, this barrier
is ``softened'' and the trajectories 
(blue curves in Fig.~\ref{trajectoriesNumerical}) 
differ a bit from the analytical approximation in 
Fig.~\ref{trajectories}.
For smaller $E'/E$ however, they converge to the analytical approximation 
(orange and green curves in Fig.~\ref{trajectoriesNumerical}) 
and already for $E'/E=10^{-3}$ 
(green curves in Fig.~\ref{trajectoriesNumerical}) 
they are virtually indistinguishable from 
the analytical approximation in Fig.~\ref{trajectories}.
They perfectly agree with the reflection picture, further justifying 
the approximations used above.

\begin{figure}
\includegraphics{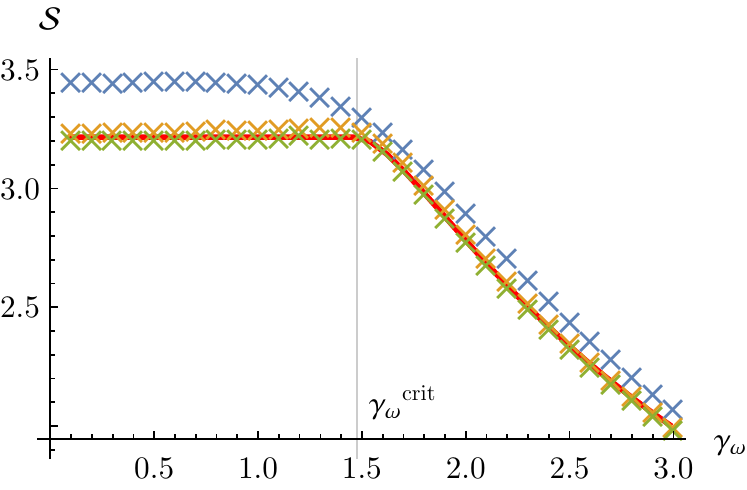}
\includegraphics{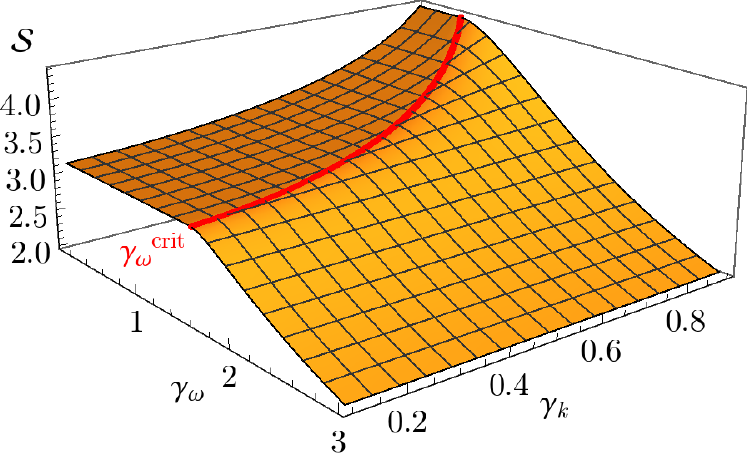}
\caption{Left: Comparison of the action~\eqref{instanton-action}
found by numerically integrating~\eqref{instanton-cosh}
for $\gamma_k=0.3$ and $E'/E=10^{-1}$ (blue), $E'/E=10^{-2}$ (orange),
and $E'/E=10^{-3}$ (green) to the integral~\eqref{action} (red line).
Right: \eqref{action} evaluated for varying $\gamma_\omega$ and
$\gamma_k$. The thick, red line visualizes the threshold
$\gamma_\omega^\text{crit}(\gamma_k)$ in \eqref{threshold}.
The action $\action$ is displayed in units of $m^2/(qE)$.
}
\label{actionPlots}
\end{figure}

With these solutions to the instanton equations, we can evaluate the
action~\eqref{instanton-action}. 
An exemplary section with $\gamma_k=0.3$ is shown in 
Figure~\ref{actionPlots}, left panel. 
While the results for $E'/E=10^{-1}$ (blue crosses) deviate a bit 
from the analytical approximation (red line) mainly because the 
total effective field is a bit larger than $E$, the agreement 
for $E'/E=10^{-3}$ (green crosses) is excellent.
Assembling all these sections for different values of $\gamma_k$
together, we obtain the landscape plot in Figure~\ref{actionPlots}, 
right panel. 
Here we plotted the analytical result, but the landscape obtained 
numerically for $E'/E=10^{-3}$, for example, would be indistinguishable. 
The red curve in Figure~\ref{actionPlots}, right panel, marks the 
threshold $\gamma_\omega^\text{crit}(\gamma_k)$ in \eqref{threshold}.

%%%%%%%%%%%%%%%%%%%%%%%%%%%%%%%%%%%%%%%%%%%%%%%%%%%%%%%%%%%%%%%%%%%%%%%%%%%%%%%
\section{Conclusions} %\& Outlook}
%%%%%%%%%%%%%%%%%%%%%%%%%%%%%%%%%%%%%%%%%%%%%%%%%%%%%%%%%%%%%%%%%%%%%%%%%%%%%%%

Via the world-line instanton technique, we derived an analytical
estimate~(\ref{action}) for the electron-positron pair creation
probability~(\ref{probability})
induced by an electric field~(\ref{Sauter}) which genuinely depends both
on space and on time.
Superimposing a strong spatial pulse by a weak temporal pulse~(\ref{Sauter}),
we found that the weak pulse is negligible for small frequencies
$\gamma_\omega\leq\gamma_\omega^{\rm crit}$ but can enhance the pair creation
probability significantly (dynamically assisted Sauter-Schwinger effect)
for larger frequencies $\gamma_\omega>\gamma_\omega^{\rm crit}$ with the
threshold~(\ref{threshold}) depending on the spatial Keldysh
parameter~(\ref{Keldysh}).
Note that we treated both, the strong $\propto E$ and the weak $\propto E'$ 
field non-perturbatively. 
This becomes evident by the reflection of the instanton trajectory in 
Fig.~\ref{trajectories}, for example, which is a drastic change and thus 
cannot be observed by a perturbative expansion around the strong field 
solution (see, e.g., \cite{catalysis}). 
Similarly, the space and time dependence is fully taken into account -- 
the threshold behavior~(\ref{threshold}) requires going beyond the locally 
constant field approximation or gradient expansion
(see, e.g., \cite{Kleinert+Xue}). 

In the homogeneous limit $\gamma_k\downarrow0$, this threshold
$\gamma_\omega^{\rm crit}$ converges to $\pi/2$ in accordance with
\cite{assisted}. 
If the spatial Keldysh parameter approaches unity $\gamma_k\uparrow1$,
on the other hand, the threshold $\gamma_\omega^{\rm crit}$ goes to zero.
In this case $\gamma_k\uparrow1$, the size of the spatial Sauter pulse is
barely enough to produce electron-positron pairs and the instanton
loop becomes very large, cf.~$x_1^*={\rm artanh}(\gamma_k)/k$ for $b=0$.
This equation $x_1^*={\rm artanh}(\gamma_k)/k$ can be rewritten as 
$m=qA_0(x_1^*)$ which shows that the potential difference between 
the spatial turning points $+x_1^*$ and $-x_1^*$ equals the energy
gap of $2mc^2$.
For $\gamma_k\uparrow1$, the total asymptotic
electrostatic potential difference approaches this energy gap and thus the 
spatial turning points $\pm x_1^*$ move to infinity. 
Quite intuitively, even comparably small frequencies (leading to poles at
large distances to the origin) can have an impact in this limit.

Note that we focused on the exponent $\action$ in the electron-positron pair 
creation probability~(\ref{probability}) given by the instanton 
action~(\ref{instanton-action}) in this work. 
The pre-factor in front of the exponential can also be derived 
(at least numerically) via the world-line instanton method by studying 
small perturbations around the instanton trajectory 
\footnote{
Note that, in contrast to the instanton action $\action$, the pre-factor
does depend on the spin, i.e., may differ for scalar and spinor QED.}%}
. 
Hence, this pre-factor is almost unaffected by the weak pulse $\propto E'$
below threshold $\gamma_\omega<\gamma_\omega^{\rm crit}$, but it is expected to 
depend on $E'$ above threshold $\gamma_\omega>\gamma_\omega^{\rm crit}$. 
Thus, for extremely small $E'$, this pre-factor could partly counteract the 
exponential enhancement of the pair creation probability. 
However, since the parameter dependence (e.g., power-law) of this pre-factor 
is sub-dominant compared to the exponential dependence on $\action$, there 
will be a large region of parameter space where the dynamically assisted 
Sauter-Schwinger effect applies 
\footnote{
If $E'$ becomes too small, the approach used here breaks down eventually.
On the one hand, the small value of $E'$ invalidates the large-$m$ expansion
(i.e., the instanton method), and, on the other hand, approximating $E'$ 
by a classical field is no longer justified.}
.  

In order to get a feeling for the orders of magnitude of the involved 
parameters, let us insert some potentially realistic example values.  
An optical laser focus with a focal field strength of $E=10^{16}~\rm V/m$ 
corresponding to an intensity of order $10^{25}~\rm W/cm^2$ has a very small 
$\gamma_k$.
Hence we get $\gamma_\omega^{\rm crit}\approx\pi/2$ corresponding to a threshold 
frequency of $\omega^{\rm crit}\approx6~\rm keV$, which is well below the energy 
gap $2mc^2\approx1~\rm MeV$ and could be reached by an x-ray free-electron 
laser (XFEL) or by higher harmonic focusing \cite{High-Harmonics}, 
for example. 
If the strong field itself is generated by such high-energy sources, 
on the other hand, the resulting $\gamma_k$ can approach unity which 
lowers $\gamma_\omega^{\rm crit}$ and thus $\omega^{\rm crit}$. 
An XFEL with 12~keV, for example, has $\gamma_k\approx1/2$ and thus we get 
$\omega^{\rm crit}\approx5~\rm keV$ while an XFEL with 24~keV corresponds to 
$\gamma_k\approx1$ such that $\omega^{\rm crit}$ vanishes. 
In these cases, the actual frequency $\omega$ exceeds the threshold 
frequency $\omega^{\rm crit}$ indicating that the dynamically assisted 
Sauter-Schwinger effect plays a role. 

However, these numbers should only be understood as an illustration because 
the profile~(\ref{Sauter}) does not portray a real laser pulse.
This motivates the study of more realistic field configurations, 
for example transverse fields such as $A_z(t,x)$. 
Preliminary investigations indicate that the world-line instanton method can 
also be applied to these more complex field profiles and that the general 
behavior of dynamically assisted Sauter-Schwinger effect persists. 
This will be the subject of future studies, see also \cite{Linder}.

%%%%%%%%%%%%%%%%%%%%%%%%%%%%%%%%%%%%%%%%%%%%%%%%%%%%%%%%%%%%%%%%%%%%%%%%%%%%%%%
\section{Outlook}
%%%%%%%%%%%%%%%%%%%%%%%%%%%%%%%%%%%%%%%%%%%%%%%%%%%%%%%%%%%%%%%%%%%%%%%%%%%%%%%

Going a step beyond the scenario considered here, we may ask the question of 
what we can learn from this analysis for more general tunneling phenomena in 
the presence of additional time-dependent fields. 
One lesson might be that the characteristic time scale of the additional 
temporal dependence required for enhancing the tunneling probability is not 
set by the height of the tunneling barrier, for example, but by the imaginary 
turning time $x_0^*$ of the associated instanton solution 
(in our case $\omega x_0^*=\pm\pi/2$).
In non-relativistic quantum mechanics, this imaginary turning time has been 
identified with the traversal time for tunneling \cite{Traversal}, 
i.e., the time the particle needs for tunneling through the barrier, 
but this identification is not undisputed.

\begin{figure}
\begin{center}
\includegraphics[width=0.4\linewidth]{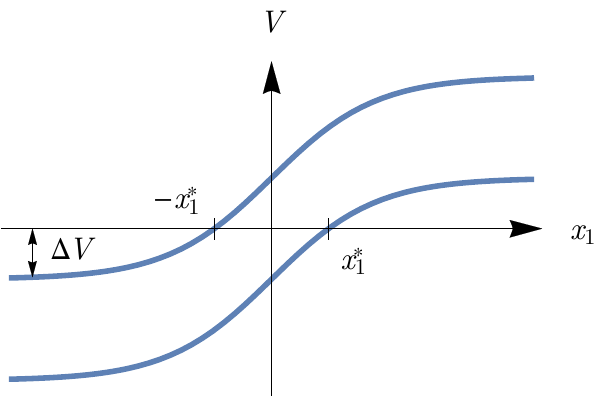}
\hspace{0.07\linewidth}
\includegraphics[width=0.4\linewidth]{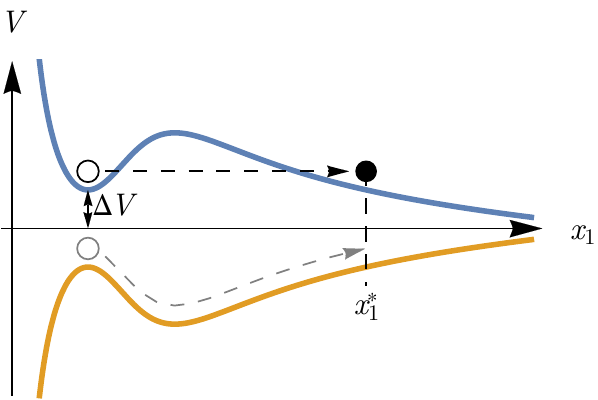}
\end{center}
\caption{
Sketch of the potential landscape $V(x)$ for the spatial Sauter 
profile $E(x)=E/\cosh^2(kx)$ (left) in comparison to non-relativistic 
tunneling out of a local minimum (right), 
as well as the inverted potential $-V(x)$ which is ``felt'' 
by the non-relativistic instanton. 
In both cases, the spatial turning points $x_1^*$ move to infinity if the 
energy gain $\Delta V$ tends to zero.}
\label{potential}
\end{figure}

Another lesson might be the relation between the final kinetic energy of the 
particle (after tunneling) and the turning points in space and (imaginary) 
time of the associated instanton solution. 
For example, let us consider a general smooth and stationary potential 
landscape $V(x)$ which contains a local minimum and vanishes at spatial 
infinity, see Fig.~\ref{potential}.
%
%%%%%%%%%%%%%%%%%%%%%%%%%%%%%%%%%%%%%%%%%%%%%%%%%%%%%%%%%%%%%%%%%%%%%%%%%%%%%%%
%%%%%%%%%%%%%%%%%%%%%%%%%%%%%%%%%%%%%%%%%%%%%%%%%%%%%%%%%%%%%%%%%%%%%%%%%%%%%%%
%
In non-relativistic quantum mechanics, the probability for tunneling from the 
local minimum out to infinity can also be estimated via an instanton solution,
which basically corresponds to the motion of the particle within the inverted 
potential landscape $-V(x)$.
Now, if the kinetic energy at infinity (i.e., $\Delta V$) becomes very small
-- which corresponds to the limit $\gamma_k\uparrow1$ in the Sauter-Schwinger 
scenario -- the spatial turning point $x_1^*$ and thus also the imaginary 
turning time $x_0^*$ become very large, suggesting that small $\omega$ can 
have a large impact. 

As a potential application, one could envisage nuclear $\alpha$-decay,
for example.  
In the usual Gamow picture \cite{Gamow}, this process can be described by 
tunneling through a potential barrier which is dominated by the Coulomb 
repulsion far away from the nucleus while the nuclear attraction prevails
at smaller distances. 
The height of this barrier is typically a few tens of MeV while its width,
i.e., the spatial turning point $x_1^*$, depends on the final kinetic energy 
of the $\alpha$-particle $E_\alpha$.

Because the instanton trajectory grows very large for small
$E_\alpha/m_\alpha$, the asymptotic behavior of the field 
(for large $x_1$) dominates since the 
``instanton spends most of its time there'', 
see also \cite{Gies+Torgrimsson}.
We can calculate both the turning points and the instanton action to 
leading order in $E_\alpha/m_\alpha$ by considering an electric field 
that asymptotically corresponds to a Coulomb potential. 
Identifying 
$\gamma_k = (1+E_\alpha/m_\alpha)^{-1}$ the spatial turning point is
\begin{equation}
    x_1^* = \frac{q_\alpha q_\text{nucleus}}{4 \pi \varepsilon_0 m_\alpha}
        \frac{1}{\gamma_k(1-\gamma_k)} \approx
        \frac{q_\alpha q_\text{nucleus}}{4 \pi \varepsilon_0 E_\alpha},
\end{equation}
which is quite large (in comparison to other nuclear length scales) for typical decays.
For example, $^{238}$U has $E_\alpha \approx 4.3$ MeV and $x_1^*\approx 62$~fm.
The imaginary turning time is given by an integral similar
to~(\ref{thresholdIntegral}):
\begin{equation}
    x_4^*
    %/{\textstyle\frac{m_\alpha}{q_\alpha E}} 
    \approx
    \frac{m_\alpha}{q_\alpha E}
    \int_0^1\frac{\mathrm{d}u}{\left(1-\gamma_k\sqrt{1-u^2}\right)^2}
    %= \frac{1}{1-\gamma_k^2}+\left(\pi-\arccos(\gamma_k)\right)
    %\frac{\gamma_k}{\left(1-\gamma_k^2\right)^{3/2}},
    \approx 
    \frac{m_\alpha}{q_\alpha E}
    \frac{\pi}{\left(2 E_\alpha / m_\alpha\right)^{3/2}},
    \label{imaginary turning time}
\end{equation}
so the critical threshold for dynamical assistance in this case is
\begin{equation}
    \omega^\text{crit} = \frac{\pi}{2 x_4^*}\approx
    \frac{\pi}{Z}\frac{\varepsilon_0 m_\alpha}{q^2}
    \left(2\frac{E_\alpha}{m_\alpha} \right)^{3/2}.
    %\frac{\left(1-\gamma_k^2\right)^{3/2}}
    %{\sqrt{1-\gamma_k^2}+\gamma_k(\pi-\arccos(\gamma_k))}.
\end{equation}
Again, for a typical decay, this is on the order of 100~keV 
(e.g., for $^{238}$U, we get $\omega^\text{crit}=$~157~keV), 
far less then the other characteristic energy scales in the 
problem, such as the $\alpha$-particle mass, its final kinetic 
energy or the barrier height $Zq^2/2\pi\varepsilon_0R_\text{nucleus}$ 
($\approx$~35~MeV for $^{238}$U) used in the usual tunneling model.

Let us also calculate the instanton action (and thus the lifetime of
the decay) analogously to~(\ref{action}). 
We have to slightly adapt the integral expression for the action in this 
case, because in the Sauter-Schwinger scenario (Fig.~\ref{potential}, left) 
there are two spatial turning points $\pm x^*_1$ moving to $\pm \infty$ as 
$\Delta V\to 0$, whereas for the $\alpha$-particle (Fig.~\ref{potential}, right) 
there is only one $x^*_1 \to \infty$. 
This amounts to a factor of $\frac{1}{2}$ in the leading-order expression 
for the instanton action:
\begin{align}
    \mathcal{S} &\approx \frac{1}{2}
    \frac{q_\alpha q_\mathrm{nucleus}}{4\pi\varepsilon_0}
    \frac{4}{\gamma_k^2} \int_0^{\frac{1}{1-\gamma_k}}\mathrm{d}u
    \sqrt{\gamma_k^2-\left(1-\frac{1}{u}\right)^2} \nonumber\\
    &\approx \frac{q_\alpha q_\mathrm{nucleus}}{2 \varepsilon_0}
    \frac{1}{\sqrt{2 E_\alpha/m_\alpha}}\nonumber\\ \label{coulombTunneling}
    &= \frac{q^2 \sqrt{m_\alpha}}{\varepsilon_0} \frac{Z}{\sqrt{E_\alpha}}. 
    %&= \underbrace{\frac{q^2 Z_\alpha}{\varepsilon_0}
    %\sqrt{\frac{m_\alpha}{2\,\text{MeV}}}}_{\approx 7.92\ \lightning\ 3.4}
    %\frac{Z}{\sqrt{Q_\alpha}},
    %\frac{1}{\gamma_k} \frac{1}{\sqrt{1-\gamma_k^2}},
\end{align}
To leading order in $E_\alpha/m_\alpha$, we reproduce the well known
Geiger-Nuttall law~\cite{Geiger+Nuttall}
\begin{equation}
    \ln(P) \propto -\mathcal{S} \propto 
    -\frac{Z}{\sqrt{E_\alpha}}.
\end{equation}
The expression~\eqref{coulombTunneling} is also obtained as the leading-order
contribution in the WKB formula for an $\alpha$-particle tunneling through
a static Coulomb potential.
Via the imaginary turning time in \eqref{imaginary turning time}, we may now 
estimate the possibility of dynamically assisting this tunneling process. 

Note that the mechanism sketched above is quite different from enhanced 
$\beta$-decay and similar ideas 
(see \cite{Becker-Scully,Baldwin+Wender,Reiss-PRC,Friar-Reiss}
for an incomplete selection) 
where the available phase space for the emitted or converted 
electrons or positrons is modified by the electromagnetic field.  

%%%%%%%%%%%%%%%%%%%%%%%%%%%%%%%%%%%%%%%%%%%%%%%%%%%%%%%%%%%%%%%%%%%%%%%%%%%%%%%
%\acknowledgments
%%%%%%%%%%%%%%%%%%%%%%%%%%%%%%%%%%%%%%%%%%%%%%%%%%%%%%%%%%%%%%%%%%%%%%%%%%%%%%%

%R.~S.~acknowledges fruitful discussions with G.~Dunne and H.~Gies
%as well as R.~Alkofer, F.~Hebenstreit, and many others.

%%%%%%%%%%%%%%%%%%%%%%%%%%%%%%%%%%%%%%%%%%%%%%%%%%%%%%%%%%%%%%%%%%%%%%%%%%%%%%%
%%%%%%%%%%%%%%%%%%%%%%%%%%%%%%%%%%%%%%%%%%%%%%%%%%%%%%%%%%%%%%%%%%%%%%%%%%%%%%%
%%%%%%%%%%%%%%%%%%%%%%%%%%%%%%%%%%%%%%%%%%%%%%%%%%%%%%%%%%%%%%%%%%%%%%%%%%%%%%%
%%%%%%%%%%%%%%%%%%%%%%%%%%%%%%%%%%%%%%%%%%%%%%%%%%%%%%%%%%%%%%%%%%%%%%%%%%%%%%%

\end{document}